\documentclass{INTERSPEECH2023}

\interspeechcameraready

\title{Behavioral Analysis of Pathological Speaker Embeddings of Patients During Oncological Treatment of Oral Cancer}
\name{Jenthe Thienpondt$^1$, Caroline M. Speksnijder$^2$, Kris Demuynck$^1$}
\address{
  $^1$IDLab, Department of Electronics and Information Systems, Ghent University - imec, Belgium\\
  $^2$Department of Oral and Maxillofacial Surgery and Special Dental Care, University Medical Center Utrecht, Utrecht University, The Netherlands}
\email{jenthe.thienpondt@ugent.be, c.m.speksnijder@umcutrecht.nl, kris.demuynck@ugent.be}

\begin{document}

\maketitle



\begin{abstract}
In this paper, we analyze the behavior of speaker embeddings of patients during oral cancer treatment. First, we found that pre- and post-treatment speaker embeddings differ significantly, notifying a substantial change in voice characteristics. However, a partial recovery to pre-operative voice traits is observed after 12 months post-operation. Secondly, the same-speaker similarity at distinct treatment stages is similar to healthy speakers, indicating that the embeddings can capture characterizing features of even severely impaired speech. Finally, a speaker verification analysis signifies a stable false positive rate and variable false negative rate when combining speech samples of different treatment stages. This indicates robustness of the embeddings towards other speakers, while still capturing the changing voice characteristics during treatment. To the best of our knowledge, this is the first analysis of speaker embeddings during oral cancer treatment of patients.

\end{abstract}

\section{Introduction}



Oral cancer is a type of cancer that can develop in various locations within the oral cavity, predominantly originating in the tissues of the mouth~\cite{cancer_globocan}. It is a serious and potentially life-threatening condition that can cause significant damage to the affected tissues and spread to other parts of the body. Common risk factors for oral cancer include tobacco usage and excessive alcohol consumption~\cite{smoking_alcohol, smoking_alcohol_2}. Treatment options for oral cancer typically include surgery, radiation therapy and chemotherapy, which may be used in isolation or in conjunction with each other, depending on the stage and location of the cancer.

In prior research, it is shown that oncological treatment of oral cancer can be accompanied with impaired speech capabilities, including articulation and intelligibility~\cite{speech_impair, speech_impair_2, speech_impair_3}. Subsequent research found reduced speech abilities even after extensive recovery periods up to 12 months after surgical intervention~\cite{speech_outcome}. Another study~\cite{path_tongue}, showed a significant decrease in tongue function during oral cancer treatment, which can potentially be an important contributor to post-intervention speech impairment. Other studies~\cite{path_asr_2008, path_asr_2022} observed a significant decrease in speech recognition transcription accuracy when comparing healthy speakers to a group of patients diagnosed with oral cancer in various treatment stages.

However, to the best of our knowledge, there is no prior research on the behavior of speaker embeddings of patients treated for oral cancer. Speaker embedding similarity, in contrast to conventional intelligibility rating systems, could provide an objective and text-independent measurement of changing voice characteristics without relying on any human perceptual evaluation of pathological speech.




In recent years, speaker verification has gained significant performance increases due to the availability of large and labeled datasets~\cite{vox1, vox2}, a significant increase in computational power and the advent of specialized deep learning models, including the x-vector architecture~\cite{x_vectors, x_vector_wide}, ECAPA-TDNN~\cite{ecapa_tdnn} and fwSE-ResNet~\cite{freq_paper}. Low-dimensional speaker embeddings can be extracted from these models and have shown to capture a wide variety of speaker characteristics, including gender, age, spoken language and emotional state~\cite{x_vector_emotion, x_vector_gender, x_vector_gender_age}.


In this paper, we want to analyze the behavior of speaker embeddings at different stages during oral cancer treatment on multiple properties. First, how do the speaker characteristics, according to the speaker embeddings, evolve between the pre- and post-intervention stages. Subsequently, we want to compare this to previous research results and establish the feasibility of potential usage of speaker embeddings during the oral cancer treatment procedure of a patient. Secondly, assess the intra-session robustness of speaker embeddings of patients based on speech samples recorded at the same session during oral cancer treatment and compare this to a cohort of non-pathological speakers. Finally, perform a speaker verification analysis when combining utterances of several steps in the intervention trajectory of the patients with the goal of analyzing the robustness of the pathological embeddings towards other speakers.





\section{Pathological speaker embeddings}
\label{s:pathological_speaker_embeddings}

\begin{table}[th]
  \caption{Dataset composition of patients with oral cancer.}
  \label{tab:dataset}
  \centering
  \begin{tabular}{lccc}
    \toprule
    \multicolumn{1}{c}{} &
    \multicolumn{1}{c}{\textbf{\# Male}} &
    \multicolumn{1}{c}{\textbf{\# Female}} &
    \multicolumn{1}{c}{\textbf{\# Total}} \\
    \midrule
    \textbf{Tumor Stage} & & & \\
    \hspace{3mm}T1 ($<$2 cm) & 5 & 3 & 8 \\
    \hspace{3mm}T2 (2-4 cm) & 11 & 6 & 17 \\
    \hspace{3mm}T3 ($>$4 cm) & 3 & 3 & 6 \\
    \hspace{3mm}T4 (metastasis) & 15 & 11 & 26 \\
    \textbf{Reconstruction Type} & & & \\
    \hspace{3mm}Primary Closure & 6 & 8 & 14 \\
    \hspace{3mm}Local Flap & 2 & 0 & 2 \\
    \hspace{3mm}Free Flap & 19 & 8 & 27 \\
    \hspace{3mm}Bone Flap & 7 & 7 & 14 \\
    \bottomrule
  \end{tabular}
  
\end{table}


The speech samples in the analysis of this paper were collected from 57 Dutch patients with primary oral carcinoma taken at the University Medical Center Utrecht (UMC Utrecht) and the Radboud University Medical Center (Radboudumc) in the Netherlands between January 2007 and August 2009. The study protocol (study ID: NL1200604106) was approved by the Ethics Committees of the UMC Utrecht and Radboudumc. All participants received written information and provided their signed informed consent. The oncological treatment of the patients consists of surgery and subsequent radiotherapy. In addition, samples were also collected from 60 healthy speakers, matched for age and gender, as the control group~\cite{healthy_speak}. Speech samples of patients were taken within 4 weeks before oncological intervention, 4 to 6 weeks after both surgery and radiotherapy and 6 and 12 months after surgery during the recovery phase. The healthy control group has speech samples only taken once. At each sampling session, two speech utterances are collected from the speakers by reading two short, phonetically diverse texts which will be referred to as \textit{text1} and \textit{text2} in this paper, respectively. The texts and recording equipment is kept consistent across all sampling sessions. The average duration of all collected speech samples is 49.6 seconds.

In addition, the tumor stage, as indicated by T of the commonly used TNM cancer staging system~\cite{tnm_staging} of the patients were also collected during the pre-intervention period. The T variable ranges from T1, indicating small tumors, to T4, indicating large tumors which have potentially invaded nearby structures, known as metastasis. Furthermore, the reconstruction type of the oral cancer surgical procedure is also collected, existing of primary closure, free flap, local flap, and bone flap reconstruction. Primary closure refers to the immediate closure of the incision after the removal of cancerous tissue. Local flap reconstruction uses adjacent oral cavity tissue to reconstruct the affected area after tumor removal, while free flap reconstruction uses tissue from another body part. Bone flap reconstruction is used to rebuild bone structures inside the oral cavity after removal of the cancer tumors. The composition of the speaker characteristics of the patients in the dataset is given in Table~\ref{tab:dataset}.


The speaker embeddings are extracted from the state-of-the-art speaker verification fwSE-ResNet34 model presented in~\cite{freq_paper}. This architecture extends the popular ResNet~\cite{resnet} backbone with a speech-adapted version of Squeeze-Excitation (SE)~\cite{se_block} and incorporates positional encodings to extend the spatial invariance of the 2D convolutional kernels with a notion of frequency positional information. The model is optimized using the Additive Angular Margin (AAM) softmax loss function~\cite{arcface}, resulting in the cosine distance being the similarity metric between speaker embeddings. More information about the architecture and training procedure can be found in the accompanying paper~\cite{freq_paper}. We note that this includes using the same training set, which solely exists of the development part of VoxCeleb2~\cite{vox2}, with no form of subsequent domain adaptation to pathological speech.


\section{Pathological speaker analysis}

It is shown that various functions related to the oral cavity are impacted by surgical and radiotherapy interventions, including the masticatory, swallowing and speech capabilities \cite{path_mastication, speech_impair_3, path_tongue}. To analyze the evolution of the speaker identifying characteristics of patients undergoing oral oncological treatment, we calculate the cosine similarities speaker-wise between the pre-operative \textit{text1} embedding and all \textit{text2} embeddings at different stages in the treatment trajectory. We also calculate the cosine similarities between the \textit{text1} and \textit{text2} embeddings of the healthy speakers to be compared to the pre-operative embedding behavior of the patients.

\begin{figure}[t]
  \centering
  \includegraphics[width=\linewidth]{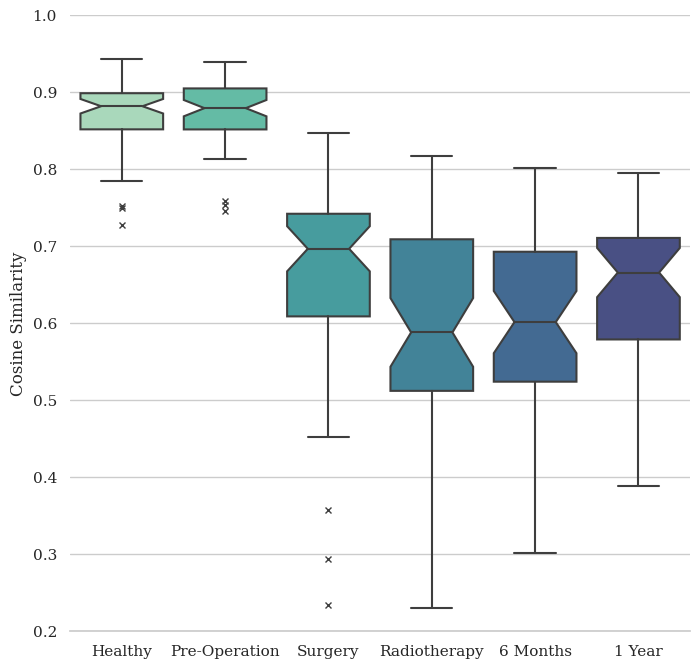}
  \caption{Tukey-style box plot depicting the evolution of pre-operative speaker embedding similarity of patients (n=57) during oral cancer treatment. Speaker similarity of a healthy control group (n=60) is included as reference. Notch width indicates the 95\% confidence interval of the median.}
  \label{fig:speech_production}
\end{figure}


Figure~\ref{fig:speech_production} depicts a box plot describing the evolution of the speaker similarity relative to the pre-operative speaker embeddings. We observe no significant difference between the pre-operative speaker similarity of the pathological group in comparison to the healthy set of speakers. While pre-operative speech impairment is usually limited for patients diagnosed with oral cancer in comparison to the post-intervention condition~\cite{speech_outcome_indian}, it is encouraging to observe similar behavior of pre-operative pathological speakers and the healthy control group. Section~\ref{ssec:intra_session} analyzes the intra-session robustness of the speaker embeddings in more detail.

A significant decrease in pre-operative speaker similarity is observed after surgical treatment of the patients. It is previously shown that surgical intervention in oral cancer treatment has a significant negative impact on a wide variety of oral function abilities, including self-reported speech capability~\cite{path_speech_self_report, speksni_retro}. Those findings are reinforced by observing a comparable degradation between the pre-operative and post-operative speaker embedding similarity, which provides an objective and robust measurement of changing voice characteristics.

Radiotherapy during oral oncological treatment can potentially impact important tissues related to speech production~\cite{radiotherapy}. However, the cumulative effect on oral function of post-operative radiotherapy strongly depends on variables such as tumor location, tumor stage and reconstruction type~\cite{path_speech_self_report}. In our results, an additional significant change in voice characteristics is discerned after the post-operative radiotherapy stage in the treatment trajectory. We also observe a substantial increase in variability between pre-operative and post-radiotherapy speaker similarity, suggesting the final extent of change in voice characteristics is highly dependent on some underlying variables.

Both an increased pre-operative speaker similarity and decreased variability is noted after the 6-month recovery period, relative to the post-radiotherapy stage, with a similar trend in the following 6 months. This indicates that voice characteristics tend to return to the pre-operative state to a certain extent for at least a 1-year period post-intervention.

\begin{figure}[t]
  \centering
  \includegraphics[width=\linewidth]{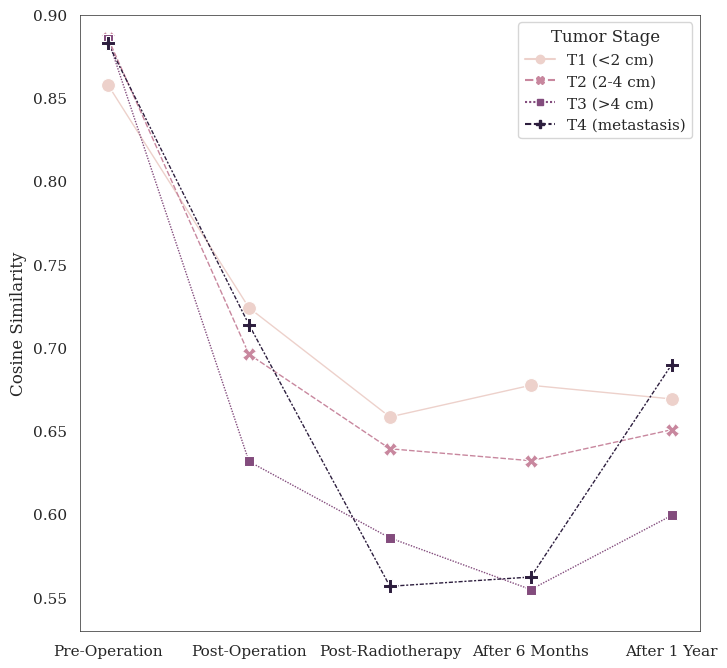}
  \caption{Effect of tumor stage, as measured by T of the TNM cancer staging model, on the evolution of pre-operative voice similarity of patients during oral cancer treatment.}
  \label{fig:tumor_stage}
\end{figure}

\subsection{Tumor stage impact on voice characteristics}

Figure \ref{fig:tumor_stage} depicts the change in pre-operative voice characteristics for each subgroup of patients based on tumor stage determined before intervention. The figure shows the mean cosine similarity between the pre-operative \textit{text1} and pre- and post-operative \textit{text2} embeddings for each subgroup. The number of speakers in each group is given in Table \ref{tab:dataset}.

We notice an inversely proportional relationship between the tumor size and the pre-operative speaker similarity at the post-intervention stages. This corroborates previous research which suggests that late-stage tumors were associated with poorer post-operative speech outcomes, including reduced speech intelligibility and decreased vocal quality~\cite{path_speech_outcomes}. Notably, this is accompanied with a more pronounced recovery towards pre-operative speaker characteristics in the T3 and T4 groups after the 1-year post-intervention period. This suggests that the additional severity of post-radiotherapy changes in speaker characteristics in the late-stage tumor groups is partially or even completely offset after sufficient recovery time.


\begin{figure}[t]
  \centering
  \includegraphics[width=\linewidth, height=7.35cm]
  {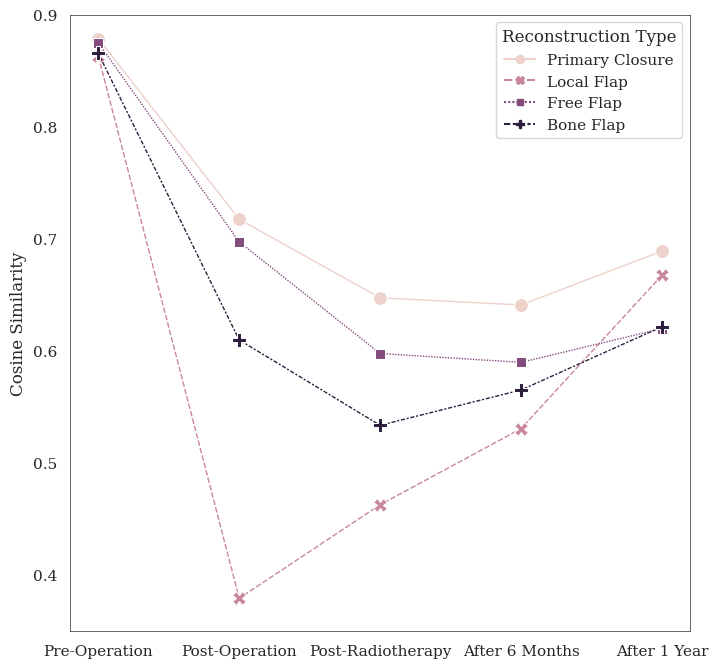}
  \caption{Effect of surgical reconstruction type on the evolution of pre-operative voice similarity of patients during oral cancer treatment.}
  \label{fig:reconstruction}
\end{figure}

\subsection{Reconstruction type impact on voice characteristics}

Likewise, Figure \ref{fig:reconstruction} shows the evolution of the pre-operative mean speaker similarity according to the type of reconstructive surgery performed. We observe that primary closure has the least significant impact on post-intervention voice characteristics in comparison to flap-based reconstruction. This supports previous research in which patients treated with primary closure were rated higher in speech intelligibility~\cite{closure_vs_flaps}. Notable is the significantly more severe change of voice characteristics of patients undergoing restorative local flap surgery in comparison to free flap surgery. This can possibly be attributed to the removal of tissue from the oral cavity during local flap surgery, as opposed to tissue removal from other parts of the body in free flap surgery. The removal of tissue in the oral cavity can potentially devise an additional degree of voice transformation in the patient in the case of local flap restoration. However, we note that the number of local flap surgeries in our dataset is limited.


\subsection{Intra-session robustness of pathological embeddings}
\label{ssec:intra_session}

State-of-the-art speaker embeddings have shown to robustly capture speaker characteristics in a variety of challenging conditions, including severe background noise, short sampling duration and language switching~\cite{score_shift}. However, it is an open question how well these embeddings can identify speakers who have had severe medical intervention in the oral cavity region. Surgery related to oral cancer treatment can have a severe impact on the structural composition of the vocal tract, which could potentially both limit or enhance the identifying characteristics captured by the speaker embeddings. In this section, we analyze the intra-session robustness of the speaker embeddings at all stages during oral cancer treatment.

\begin{figure}[t]
  \centering
  \includegraphics[width=\linewidth]{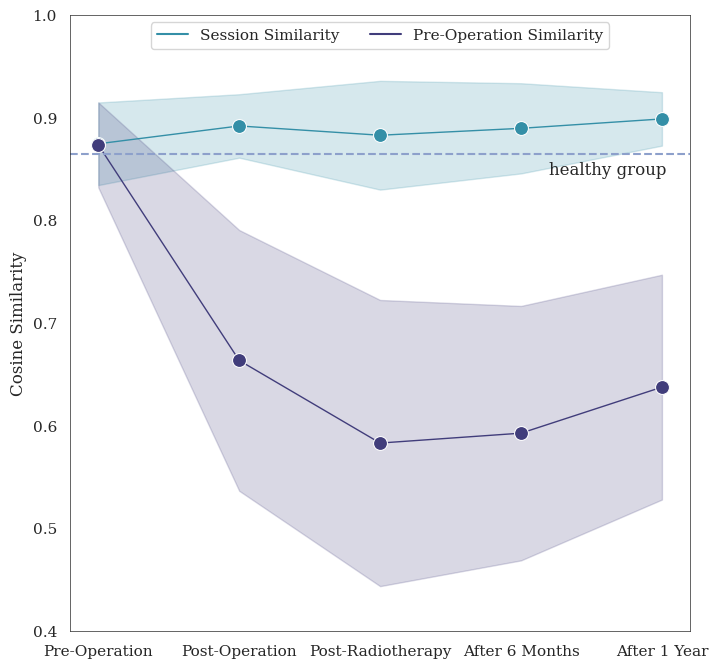}
  \caption{Intra-session and inter-session (relative to the pre-operative session) same-speaker similarity of patients during oral cancer treatment. The dotted line indicates the mean same-speaker similarity of the healthy control group.}
  \label{fig:intra_session}
\end{figure}


To establish the intra-session robustness of the speaker embeddings, we calculate the cosine similarity between the \textit{text1} and \textit{text2} embedding of each patient at all sampling sessions during the treatment trajectory. The session-wise mean and standard deviation of the same-speaker cosine similarities is shown in Figure~\ref{fig:intra_session}. As a reference, the mean similarity between the embeddings from the same speakers in the healthy group is indicated by the dotted line. For comparison, we also plotted the mean and standard deviation of the speaker-wise similarities between the pre-operative \textit{text1} and post-intervention \textit{text2} embeddings.

We can observe that the mean intra-session similarity is very consistent during the complete oral cancer treatment trajectory, even slightly exceeding the healthy control group. This indicates that the speaker embeddings are able to capture robust and distinguishing voice characteristics of speakers, even after substantial oncological intervention in the oral cavity, given the changed voice characteristics are temporally stable. This is notable due to the training set of the speaker embedding extractor not containing any comparable pathological speakers. This implies no domain-specific adaption of the training procedure of the speaker embedding extractor is needed, which greatly alleviates the potential medical usage of speaker embeddings in oral cancer treatment.

\subsection{Pathological speaker verification analysis}
In this section we want to analyze the behavior of pathological speaker embeddings in a speaker verification setting. Speaker verification attempts to solve the task if two utterances are spoken by the same person. We create three groups of speaker verification trials based on speech samples from patients: pre-operative, pre-operative combined with post-operative and pre-operative combined with post-radiotherapy utterances. To increase the number of trials, we create consecutive, non-overlapping crops of 5 seconds of each utterance and subsequently extract the speaker embeddings as described in Section~\ref{s:pathological_speaker_embeddings}. Each trial consists of a \textit{text1} embedding paired with a \textit{text2} embedding for text-independency and we balance the amount of positive and negative trials. Results are reported using the equal error rate (EER) and a breakdown of the false positive rate (FPR) and false negative rate (FNR).

\begin{figure}[t]
  \centering
  \includegraphics[width=\linewidth,height=7.7cm]
  {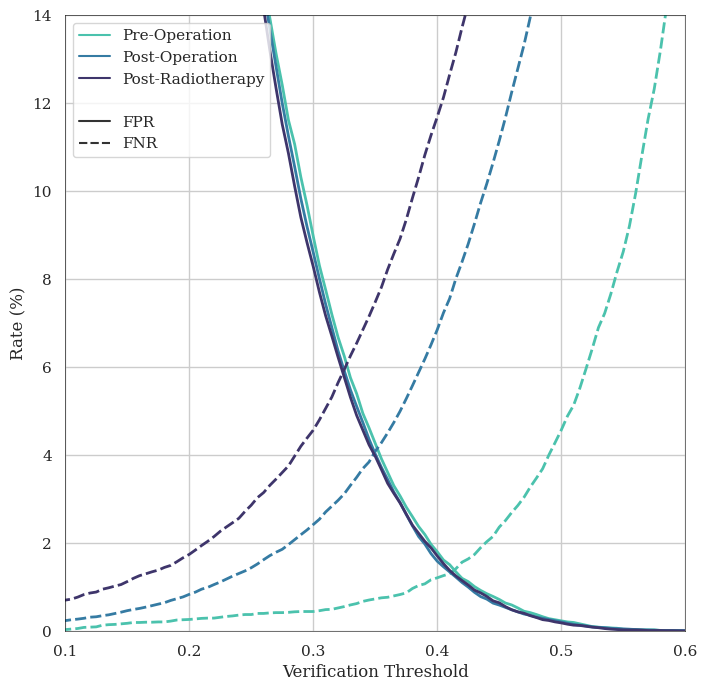}
  \caption{False positive rates (FPR) and false negative rates (FNR) of speaker verification trials consisting of pre-operative, pre-operative combined with post-operative and pre-operative combined with post-radiotherapy speech samples.}
  \label{fig:verification}
\end{figure}

\begin{table}[th]
  \caption{Speaker verification results of oral cancer patients. FPR and FNR are based on a threshold value of 0.35.}
  \label{tab:eer}
  \centering
  \begin{tabular}{lccc}
    \toprule
    \multicolumn{1}{c}{} &
    \multicolumn{1}{c}{\textbf{EER (\%)}} &
    \multicolumn{1}{c}{\textbf{FPR (\%)}} &
    \multicolumn{1}{c}{\textbf{FNR (\%)}} \\
    \midrule
    Pre-operation & 1.39 & 4.26 & 0.73 \\
    Post-operation & 4.06 & 4.03 & 4.08 \\
    Post-radiotherapy & 5.88 & 3.97 & 7.48 \\
    \bottomrule
  \end{tabular}
  
\end{table}


As shown in Table \ref{tab:eer}, the overall EER sharply increases by the subsequent addition of post-operative and post-radiotherapy samples. However, as Figure~\ref{fig:verification} indicates, the FPR of all trial groups remains almost identical, independent of the chosen speaker verification threshold. The degradation of EER can exclusively be attributed by an increase in FNR in the groups combining pre-operative and post-intervention embeddings. The implications of a stable FPR and variable FNR are desirable from an oral cancer treatment viewpoint. A stable FPR signifies a robust behavior of the speaker embeddings towards other speakers, while simultaneously still being able to capture the change in voice characteristics of the same speaker during the treatment trajectory.

\section{Future work}
As shown in this paper, the use of speaker embeddings has the potential to improve our understanding of changing voice characteristics during oral cancer treatment. Using speaker embeddings to analyze individual treatment trajectories proves viable due to a combination of intra-session robustness, objective and text-independent metrics for changing voice characteristics and no reliance on human perceptual evaluation in the process. In future work, we will attempt to investigate the feasibility of using speaker embeddings to identify potential complications or challenges that may arise during the recovery process.

\section{Conclusion}
In this paper, we analyzed the behavior of speaker embeddings of patients diagnosed with oral cancer at different stages during oncological treatment. First, we found that pre-operative and post-intervention speaker similarity significantly diminishes. However, we observe an evolution of the voice characteristics towards the pre-operative stage in the following 12-month post-operative period. Secondly, we establish the intra-session robustness of current state-of-the-art speaker embeddings on speakers with oral cancer treatment. This indicates that the embeddings can successfully capture pathological speaker characteristics, given the pathology is temporally stable. Finally, we observe a stable false positive rate and variable false negative rate in a speaker verification analysis when speech samples are used from different stages in oral cancer treatment. This signifies a stable behavior of the embeddings towards other speakers while still being able to capture the change in voice characteristics during oral oncological treatment.

\bibliographystyle{IEEEtran}
\bibliography{mybib}

\end{document}